\documentclass[manuscript]{acmart}
\usepackage[shortlabels]{enumitem}
\usepackage{tabularx}
\usepackage{xcolor}

\AtBeginDocument{%
  }

\copyrightyear{2024}
\acmYear{2024}
\setcopyright{acmlicensed}
\acmConference[MUM '24]{International Conference on Mobile and Ubiquitous Multimedia}{December 1--4, 2024}{Stockholm, Sweden}
\acmBooktitle{International Conference on Mobile and Ubiquitous Multimedia (MUM '24), December 1--4, 2024, Stockholm, Sweden}
\acmDOI{10.1145/3701571.3701608}
\acmISBN{979-8-4007-1283-8/24/12}

\begin{document}

%%%%%%%%%%%%%%%%%%%
%%    "Title"    %%
%%%%%%%%%%%%%%%%%%%
\title{Designing AI Personalities: Enhancing Human-Agent Interaction Through Thoughtful Persona Design}
%%%%% Title samples
% Desiging AI Characters 
% Crafting Digital Personalities

%%%%% Subtitle samples
% Customizing Agent Characteristics for Enhanced User Interaction
% Tailoring AI Personas for Context-Specific Applications
% Balancing Identity, Culture, and Context in AI Assistants

%%%%%%%%%%%%%%%%%%%
%%   "Authors"   %%
%%%%%%%%%%%%%%%%%%%
\author{Nima Zargham}
\email{zargham@uni-bremen.de}
\orcid{0000-0003-4116-0601}
\affiliation{
  \institution{Digital Media Lab, University of Bremen}
  \country{Germany}
}

\author{Mateusz Dubiel}
\orcid{0000-0001-8250-3370}
\affiliation{%
  \institution{University of Luxembourg}
  \country{Luxembourg}}
\email{mateusz.dubiel@uni.lu}

\author{Smit Desai}
\orcid{0000-0001-6983-8838}
  \affiliation{%
  \institution{Northeastern University}
  \country{USA}}
\email{sm.desai@northeastern.edu}

\author{Thomas Mildner}
\email{mildner@uni-bremen.de}
\orcid{0000-0002-1712-0741}
\affiliation{
    \institution{Digital Media Lab, University of Bremen}
    \country{Germany}
}

\author{Hanz-Joachim Belz}
\email{hans-joachim.belz@anstrengungslos.de}
\orcid{0009-0001-0257-3078}
\affiliation{
    \institution{Freelancer}
    \country{Germany}
}

%%%%%%%%%%%%%%%%%%%%%%%%%
\renewcommand{\shortauthors}{Zargham et al.}

%%%%%%%%%%%%%%%%%%%%%%%%%%%%%
%%%%%%    "Abstract"    %%%%%
%%%%%%%%%%%%%%%%%%%%%%%%%%%%%
\begin{abstract}
In the rapidly evolving field of artificial intelligence (AI) agents, designing the agent's characteristics is crucial for shaping user experience. This workshop aims to establish a research community focused on AI agent persona design for various contexts, such as in-car assistants, educational tools, and smart home environments. We will explore critical aspects of persona design, such as voice, embodiment, and demographics, and their impact on user satisfaction and engagement. Through discussions and hands-on activities, we aim to propose practices and standards that enhance the ecological validity of agent personas. Topics include the design of conversational interfaces, the influence of agent personas on user experience, and approaches for creating contextually appropriate AI agents. This workshop will provide a platform for building a community dedicated to developing AI agent personas that better fit diverse, everyday interactions.
\end{abstract}

%%%%%%%%%%%%%%%%%%%%%%%%%%%%%%
%%%%%%%       "CCS"      %%%%%
%%%%%%%%%%%%%%%%%%%%%%%%%%%%%%


%%%%%%%%%%%%%%%%%%%%%%%%%%%%%%
%%%%%%%    "Keywords"    %%%%%
%%%%%%%%%%%%%%%%%%%%%%%%%%%%%%


%%%%%%%%%%%%%%%%%%%%%%%
%%%  Teaser Figure  %%%
%\centering
 %\includegraphics[width=0.99\columnwidth]{figures/.png}
 %\caption{}
 %\Description{}
 %\label{fig:teaser}

%%%%%%%%%%%%%
\maketitle

%%%%%%%%%%%%%%%%%%%%%%%%%%%%%%
%%%%%  Theme and Goals   %%%%%
%%%%%%%%%%%%%%%%%%%%%%%%%%%%%%
\section{Theme and Goals}
%%%% Importance of Presentation
As artificial intelligence (AI) agents become increasingly integrated into interfaces in a dialogue-based manner via both text and voice, their role in human-computer interaction (HCI) has become more crucial~\cite{clark2019state}. 
%%%%  Terminologies
Virtual agents (VAs) are software programs designed to assist users by completing tasks, answering queries, or providing information~\cite{Wang2021Examining}. 
%%%% AI Agents
AI agents are a subset of virtual assistants that employ AI techniques to understand, learn from, and adapt to user preferences over time~\cite{ruan2023tptu}. They can range from simple rule-based systems to complex, learning-based models, depending on the tasks they are designed to perform. Examples of AI agents include Siri, Google Assistant, Alexa, and ChatGPT.
%%%%% CUIs
The interfaces that allow users to interact with an agent using natural language through text or voice are referred to as Conversational User Interfaces (CUIs)~\cite{Moore2019Conversational}. 
%%%% Popularity
The growing popularity of these interfaces is largely driven by advances in natural language processing capabilities~\cite{voiceCraft2024}. While much of the research in this area has focused on improving system efficiency and accuracy~\citep{Allison2018, ZarghamDissertation2024}, it is equally important to consider the appearance and representation of AI agents. Mismatch in user expectations and agent's presentation could result in infrequent use and eventually un-use \cite{Desai_Twidale_2023, Desai_Twidale_2022, Trajkova_Martin-Hammond_2020}. Yuksel et al. suggest that an agent's visual appeal may be more critical to user satisfaction than its reliability~\cite{Yuksel:2017}. Similarly, Lopatovska et al.~\citep{Lopatovska:2019} emphasize that, depending on the task, a positive user experience (UX) may take precedence over the accuracy of outcomes. Additionally, Desai et al. \cite{Desai_Dubiel_Leiva_2024} found that the choice of persona also affects the perceived trust, likeability, and intention to adopt. This highlights that users' engagement, satisfaction, and even trust are influenced not only by the technical performance of AI systems but also by the broader interaction process, including the agent's persona and presentation.

%%%%%%%
There is a common misconception that the representation of AI agents is a superficial aspect, seemingly detached from the core components that significantly contribute to the system's effectiveness and user acceptance~\citep{khan2009attractiveness, Yuksel:2017}. However, this view is shifting. With the proliferation of AI agents, particularly in homes and smartphones, research into the persona and representation dimension of such agents has been expanding. Studies suggest that many conversational agents fail to meet user expectations as effective interlocutors \citep{jentsch2019talking, Luger:2016, murad2019don, Doyle2019Mapping, dubiel2018towards, Zargham2023FaceIt}. A potential contributing factor to this shortfall is the insufficient emphasis placed on the representation of these entities as engaging conversation partners.
%%%%%%%%
Agents like Amazon Alexa and Google Home serve complex roles, from financial advisors to health coaches, facilitating tasks that significantly impact users' daily lives~\cite{dubiel2022conversational, DesaiChin2023, Desai_Dubiel_Leiva_2024}. 
%%%%%% Limited Research
Despite rapid advancements in speech technology and its widespread commercial deployment, research into the design of these agent and their implications for users remains limited \cite{schmitt2021voice}. This scarcity of studies contributes to inconsistencies in the development and evaluation of system personas \cite{DesaiChin2021, MotalebiChoSundarAbdullah2019}. These personas are often assessed through interactive user studies, conducted both online and in laboratory settings, yet lack uniform standards and methodologies.
%%%% GPT
Moreover, with the growing popularity of ChatGPT, users can choose personas to interact with, including those modeled after celebrities like Tom Brady and Snoop Dogg \cite{Introducing2023}. Additionally, they can craft personalized personas using prompt engineering techniques. These design choices raise ethical concerns, particularly due to their potential to reinforce and perpetuate stereotypes \cite{10.1145/3643834.3661555}.  

%%%% Our Approach
To address these challenges, this workshop will convene interdisciplinary experts to refine how AI agent personas are crafted. The goal is to enhance design practices and increase the ecological validity of these systems. Throughout the workshop, through interactive demonstrations, vibrant group discussions, and hands-on prototyping activities, we will explore key questions tailored to the creation of effective and engaging personas, such as:

\begin{itemize}
    \item \textit{How can we design appropriate AI agent personas based on the context of use?}
    \item \textit{What ethical and sociotechnical challenges arise from the design choices of AI agent personas??}
\end{itemize}

We encourage collaboration between academia and industry and welcome participation from anyone interested in conversational user interface design, development, and evaluation. By bringing the interdisciplinary community together, we aim to provide insights on how to improve the development of AI agent personas tailored to specific applications and ensure higher ecological validity in their task-based evaluation.

%%%%%%%%%%%%%%%%%%%%%%%%%%%%
%%%%%%%  Organizers   %%%%%%
%%%%%%%%%%%%%%%%%%%%%%%%%%%%
\section{Organisers}
This workshop is organized by a team of experienced researchers with expertise spanning diverse fields aligned with our objectives. 
The organizers have a strong record of publications in areas such as conversational user interfaces~\cite{zargham2021MultiAgent, zargham2022Proactive, dubiel2020persuasive, dubiel2019inquisitive, dubiel2024hey, reicherts2021MayI,mildner_rules_2022, mildner_listening_2024}, human-robot interaction~\cite{McMillan2023}, multi-modal communication~\cite{bonfert2021evaluation, Avanesi2023Multimodal}, and agent persona design~\cite{zargham2021Customization}. By bringing together these disciplines, the workshop aims to foster collaboration among participants from various backgrounds, bridging gaps between these research areas and promoting transdisciplinary innovation.\\
Multiple successful workshops were held between the organizers at HCI conferences, such as HRI'23~\cite{McMillan2023}, CUI'23~\cite{Avanesi2023Multimodal}, CUI'24~\cite{voiceCraft2024}, and CHI'24~\cite{gray_mobilizing_2024, Desai2024CUICHI}, to give important topics a platform for discourse, including conversational user interfaces, ethical design, and dark patterns.\\

%%%%%%%%%%%%
\noindent \textbf{Nima Zargham} is a postdoctoral researcher in the Digital Media Lab at the University of Bremen. His research focuses on human-centered approaches for designing speech-based systems that elicit desirable user experiences. Nima has previously organized CUI-related workshops at notable conferences such as ACM/IEEE HRI 2023, ACM CUI 2023, ACM CUI 2024, and ACM CHI 24. Additionally, he served as a local chair at the ACM CHI-PLAY 2022 conference. His research efforts have resulted in publications featured in prestigious HCI venues, including CHI, CUI, and CHI-PLAY. \\

%%%%%%%%%%%%
\noindent \textbf{Mateusz Dubiel} is a research associate in the Department of Computer Science at the University of Luxembourg, where he works on developing and evaluating conversational agents. Specifically, his current research focuses on assessing the cognitive and usability implications of interfaces that feature speech and exploring their potential to inspire positive behavioral change in users. He served as Short Papers Chair for CUI '22 and was one of the General Chairs for CUI '24. \\

%%%%%%%%%%%%
\noindent \textbf{Smit Desai} is a postdoctoral researcher in the College of Art, Media and Design at the Northeastern University, Boston. His primary research focus centers around comprehending the mental models of users as they engage with conversational agents, utilizing innovative research techniques such as metaphor analysis. He leverages this valuable insight to advance the development of conversational agents in diverse social roles, including educators and storytellers. His research has yielded publications in esteemed HCI forums like CHI, TOCHI, CSCW, and CUI. \\

%%%%%%%%%%%%
\noindent \textbf{Thomas Mildner} is a postdoctoral researcher at the Digital Media Lab at the University of Bremen. His research focuses on ethical and responsible design and online wellbeing with studies exploring so-called dark patterns in social media as well as conversational technologies. To this end, Thomas collaborated to develop an ontology for dark patterns~\cite{gray_ontology_2024}. His research is published in venues including CHI, DIS, and CUI. \\

%%%%%%%%%%%%
\noindent \textbf{Hans-Joachim Belz} is a freelance user researcher and designer with over 30 years of experience designing, implementing, and managing digital products across various industries. From 2014 to 2024, he held a teaching position in ``Mobile Commerce'' at DHBW Mannheim (Baden-Wuerttemberg Cooperative State University). In 2020, his focus shifted to voice automation. In addition to his consultancy work in conversational AI, he employs Ethnomethodology and Conversation Analysis (EMCA) to study the expanding conversational capabilities of voice and multimodal user interfaces.

%%%%%%%%%%%%%%%%%%%%%%%%%%%%%%%%%%%%%
%%%%%  Schedule and Activities  %%%%%
%%%%%%%%%%%%%%%%%%%%%%%%%%%%%%%%%%%%%
\section{Format and Advertisement}
%%%% Invited Groups
This workshop seeks to connect individuals from academia and industry to engage in cross-disciplinary dialogue. We invite participants from different Human-Computer Interaction (HCI) communities, including but not limited to human-agent interaction, human-robot interaction, conversational user interfaces, and user experience design.
%%% Advertisement
To reach potential participants, we will issue a Call for Participation through several channels, including popular social media platforms (e.g., X, Facebook, LinkedIn) and relevant mailing lists. Additionally, we will directly invite researchers who have published in related fields within HCI.

%%%% Format and Requirement
The workshop will be held exclusively in person. Based on previous experiences, we expect to host between 7 to 15 attendees and receive 4 to 6 position papers. A dedicated workshop \href{https://thomasmildner.github.io/DAIP}{website} will be established to provide information about the call, key dates, and other details for prospective participants. Accepted submissions will be posted on this site before the workshop and archived afterwards.
Applicants are invited to submit position papers of 2--4 pages, presenting their work or discussing innovative ideas related to AI agent persona design and evaluation. We encourage submissions on research in progress, preliminary results for community discussion, methodology proposals, and insights gained from designing AI agents for users. All submissions will be reviewed independently by at least two workshop organizers before acceptance. We also welcome expressions of interest from those who may not have formal papers but wish to contribute to the workshop. Statements of interest should highlight relevant experience, perspectives, or ideas related to AI persona design.

%%%%%%%%%%%%%%%%%%%%%%%%%%%%%%%%%%%%%
%%%%%  Schedule and Activities  %%%%%
%%%%%%%%%%%%%%%%%%%%%%%%%%%%%%%%%%%%%
\section{Schedule and Workshop Activities}
%%%%%%
The tentative schedule and workshop activities are outlined below. Given our workshop's focus on synthesizing insights from experts across various communities and translating them into practical recommendations for persona design and evaluation, we will allocate ample time for prototyping activities, discussions, and participant interactions.

\begin{itemize}
    \item \textbf{Welcome and Introduction (30 mins):} We will begin with brief introductions from the organizers and participants, outlining the workshop's objectives and providing an overview of the day's agenda.
    \item \textbf{Keynote and Discussion (60 mins):} Our keynote address will be delivered by \textbf{Dr. Hannah Pelikan}. She is a PostDoc in the Language, Culture, and Interaction division at Linköping University, working on the WASP-HS AI in Motion project with professors Barry Brown and Mathias Broth. Her work is located at the intersection of conversation analysis and interaction design.
    %and was honored with two best paper awards (ACM/IEEE HRI, ACM CSCW) and one ACM best paper nomination (HRI). 
    Her PhD thesis on sound design for robots was selected for presentation at the HRI Pioneers workshop, which gathers the world's top student researchers in Human-Robot Interaction. 
    %Hannah is engaged in multiple international research projects and has received funding from the Swedish Innovation Agency for her collaboration with Cornell University and Responsible AI UK with the University of Nottingham. 
    Hannah regularly speaks at popular science events such as Pint of Science and Forskarfredag, and her work has been featured in several press releases, including the Cornell  Chronicle, Elektroniktidningen, and the University of Nottingham’s 100 Ways to Change the World.
    \item \textbf{Coffee Break (15 mins)}
    \item \textbf{Introduction \& Presentations (45 mins):} Each participant will have 3-5 minutes to introduce themselves and present their accepted paper. This will allow everyone to share their background, research focus, and key insights from their work.
    \item \textbf{Expressive AI Agent Persona Demonstration (45 mins):} The organizers will present a selection of AI agents, showcasing their personas and the design choices that shaped them. This presentation will set the stage for an open discussion, where participants can share their perspectives, critique the designs, and explore the implications of these choices on user experience and interaction.
    \item \textbf{Lunch (60 mins)}
    \item \textbf{Prototype activity (90 mins):} Participants will be divided into two or three small groups, each guided by a designated facilitator. Each group will be assigned a specific domain and context, such as designing an AI assistant for financial guidance or creating an agent as a social companion. The groups will start by identifying and listing the characteristics of their target audience, including their attitudes, needs, goals, and expectations. Following this, the groups will collaboratively design an agent persona, considering factors like voice, embodiment, and demographic characteristics. At the end of the session, each group will present their designed persona, discuss the rationale behind their design choices, and receive feedback from the broader audience.
    \item \textbf{Coffee Break (15 mins)}
    \item \textbf{Closing (45 mins):} The workshop will conclude with the organizers summarizing key insights from the sessions. This will be followed by a collaborative reflection, where both participants and organizers will work together to identify concrete research topics and potential outcomes. This session will also explore future research directions and strategies for upcoming venues and workshops, incorporating participants' feedback and perspectives. 
\end{itemize}

%%%%%%%%%%%%%%%%%%%%%%%%%%%%%%
%%%%%  Publishing Stuff  %%%%%
%%%%%%%%%%%%%%%%%%%%%%%%%%%%%%
\section{Plans to Publish Workshop Material}
% Community Building and Networking
The primary goal of this workshop is to promote collaboration among individuals from diverse backgrounds and perspectives who are interested in designing and evaluating AI agent personalities.
% Publication of Contributions
All accepted submissions from participants will be published on our workshop's website to ensure broad dissemination of the findings.
% Proceedings and Archival
A selection of the accepted papers will be included in the workshop proceedings, which we plan to publish on platforms such as \textit{arXiv} \footnote{https://arxiv.org/} or \textit{ceur-ws} \footnote{https://ceur-ws.org/}.
% Refined Publication
Moreover, we aim to synthesize the workshop's results into a refined publication for a relevant venue, such as CUI 2025\footnote{https://cui.acm.org}.
% Open Sharing and Future Collaboration
Key ideas and discussion points will be documented on an open online platform during and after the workshop. Participants will be encouraged to engage in future collaborative projects that arise from the discussions.
%Encouragement for Extended Submissions
% We will encourage participants to submit extended versions of their contributions for consideration in a special journal issue, which we are committed to facilitating.

%The workshop outcomes will be presented as a report and posted on the workshop's website to reach a broader audience. The organisers will use social media platforms (Mastodon and X (formerly Twitter)) to disseminate these materials.
%%%%%%%%%%%%%%%%%%%%%%%%%%%%%%%%%%%
%%%%  Call For Participation   %%%%
%%%%%%%%%%%%%%%%%%%%%%%%%%%%%%%%%%%
\section{Call For Participation}
In the rapidly evolving field of artificial intelligence agents, the design and representation of AI agent personas are crucial to shaping user interactions and experiences. Our workshop aims to bring together experts from diverse disciplines to explore and discuss the design of AI agent personas across various contexts, such as in-car assistants, educational support tools, and smart home environments.

Join us for a full-day in-person workshop to advance the discourse on AI gent personas and representations. We invite participants from both academia and industry to engage in critical discussions about conversational user interfaces (CUIs) and the development of design guidelines to enhance interaction quality.

We encourage the submission of position papers (2--4 pages in ACM single-column format, including references) that present research findings, innovative ideas, and work-in-progress related to AI agent persona design. Topics of interest include, but are not limited to:

\begin{itemize}
    \item AI Agent Representations
    \item Embodied Conversational Agents
    \item Ethical Dimension in CUI Persona Design
    \item Multimodal AI Agents
    \item Persona Prompting 
\end{itemize}

All submissions will undergo review by at least two workshop organizers. Papers will be selected based on their quality, relevance to the workshop topics, and diversity of perspectives. Please note that at least one author of each accepted submission must attend the workshop. Key details can be found below:\newline

\begin{itemize}
    \item Submission deadline: 7th of November, 2024 
    \item Acceptance Notification: 18th of November, 2024
    \item Send position papers to: \textbf{cui.research.hub@gmail.com}
    \item Website: \href{https://thomasmildner.github.io/DAIP}{https://thomasmildner.github.io/DAIP}
    \item Date and time: 1st of December, 9:00 a.m. - 16:00 p.m. (UTC+2)
\end{itemize}

{\noindent}

%%%%%%%%%%%%%%%%%%%%%%%%%%%%%
%%%%%%%  Acknowledge   %%%%%%
%%%%%%%%%%%%%%%%%%%%%%%%%%%%%
\begin{acks}
This work was partially funded by Klaus Tschira Foundation, the FET-Open Project 951846 ``MUHAI -- Meaning and Understanding for Human-centric AI'' funded by the EU program Horizon 2020, as well as the German Research Foundation DFG as part of Collaborative Research Center (Sonderforschungsbereich) 1320 ``EASE -- Everyday Activity Science and Engineering'', University of Bremen (\url{http://www.ease-crc.org/}). This work is in part supported by European Innovation Council Pathfinder program (SYMBIOTIK project, grant 101071147).
\end{acks}

%%%%%%%%%%%%%%%%%%%%%%%%%%%%%
%% the bibliography file  %%%
%%%%%%%%%%%%%%%%%%%%%%%%%%%%%
\bibliographystyle{ACM-Reference-Format}
\bibliography{sample-base}


\begin{thebibliography}{42}

%%% the Web address.

\ifx \showCODEN    \undefined \def \showCODEN     #1{\unskip}     \fi
\ifx \showDOI      \undefined \def \showDOI       #1{#1}\fi
\ifx \showISBNx    \undefined \def \showISBNx     #1{\unskip}     \fi
\ifx \showISBNxiii \undefined \def \showISBNxiii  #1{\unskip}     \fi
\ifx \showISSN     \undefined \def \showISSN      #1{\unskip}     \fi
\ifx \showLCCN     \undefined \def \showLCCN      #1{\unskip}     \fi
\ifx \shownote     \undefined \def \shownote      #1{#1}          \fi
\ifx \showarticletitle \undefined \def \showarticletitle #1{#1}   \fi
\ifx \showURL      \undefined \def \showURL       {\relax}        \fi
% invisible to TeX
\providecommand\bibfield[2]{#2}
\providecommand\bibinfo[2]{#2}
\providecommand\natexlab[1]{#1}
\providecommand\showeprint[2][]{arXiv:#2}

\bibitem[Allison et~al\mbox{.}(2018)]%
        {Allison2018}
\bibfield{author}{\bibinfo{person}{Fraser Allison}, \bibinfo{person}{Marcus Carter}, \bibinfo{person}{Martin Gibbs}, {and} \bibinfo{person}{Wally Smith}.} \bibinfo{year}{2018}\natexlab{}.
\newblock \showarticletitle{Design Patterns for Voice Interaction in Games}. In \bibinfo{booktitle}{\emph{Proceedings of the 2018 Annual Symposium on Computer-Human Interaction in Play}} (Melbourne, VIC, Australia) \emph{(\bibinfo{series}{CHI PLAY ’18})}. \bibinfo{publisher}{Association for Computing Machinery}, \bibinfo{address}{New York, NY, USA}, \bibinfo{pages}{5–17}.
\newblock
\showISBNx{9781450356244}
\urldef\tempurl%
\url{https://doi.org/10.1145/3242671.3242712}
\showDOI{\tempurl}


\bibitem[Avanesi et~al\mbox{.}(2023)]%
        {Avanesi2023Multimodal}
\bibfield{author}{\bibinfo{person}{Vino Avanesi}, \bibinfo{person}{Johanna Rockstroh}, \bibinfo{person}{Thomas Mildner}, \bibinfo{person}{Nima Zargham}, \bibinfo{person}{Leon Reicherts}, \bibinfo{person}{Maximilian~A. Friehs}, \bibinfo{person}{Dimosthenis Kontogiorgos}, \bibinfo{person}{Nina Wenig}, {and} \bibinfo{person}{Rainer Malaka}.} \bibinfo{year}{2023}\natexlab{}.
\newblock \showarticletitle{From C-3PO to HAL: Opening The Discourse About The Dark Side of Multi-Modal Social Agents}. In \bibinfo{booktitle}{\emph{Proceedings of the 5th International Conference on Conversational User Interfaces}} (Eindhoven, Netherlands) \emph{(\bibinfo{series}{CUI '23})}. \bibinfo{publisher}{Association for Computing Machinery}, \bibinfo{address}{New York, NY, USA}, Article \bibinfo{articleno}{62}, \bibinfo{numpages}{7}~pages.
\newblock
\showISBNx{9798400700149}
\urldef\tempurl%
\url{https://doi.org/10.1145/3571884.3597441}
\showDOI{\tempurl}


\bibitem[Bonfert et~al\mbox{.}(2021)]%
        {bonfert2021evaluation}
\bibfield{author}{\bibinfo{person}{Michael Bonfert}, \bibinfo{person}{Nima Zargham}, \bibinfo{person}{Florian Saade}, \bibinfo{person}{Robert Porzel}, {and} \bibinfo{person}{Rainer Malaka}.} \bibinfo{year}{2021}\natexlab{}.
\newblock \showarticletitle{An Evaluation of Visual Embodiment for Voice Assistants on Smart Displays}. In \bibinfo{booktitle}{\emph{CUI 2021-3rd Conference on Conversational User Interfaces}}. \bibinfo{publisher}{ACM}, \bibinfo{address}{New York, NY, USA}, \bibinfo{pages}{1--11}.
\newblock


\bibitem[Clark et~al\mbox{.}(2019)]%
        {clark2019state}
\bibfield{author}{\bibinfo{person}{Leigh Clark}, \bibinfo{person}{Philip Doyle}, \bibinfo{person}{Diego Garaialde}, \bibinfo{person}{Emer Gilmartin}, \bibinfo{person}{Stephan Schl{\"o}gl}, \bibinfo{person}{Jens Edlund}, \bibinfo{person}{Matthew Aylett}, \bibinfo{person}{Jo{\~a}o Cabral}, \bibinfo{person}{Cosmin Munteanu}, \bibinfo{person}{Justin Edwards}, {et~al\mbox{.}}} \bibinfo{year}{2019}\natexlab{}.
\newblock \showarticletitle{The state of speech in HCI: Trends, themes and challenges}.
\newblock \bibinfo{journal}{\emph{Interacting with computers}} \bibinfo{volume}{31}, \bibinfo{number}{4} (\bibinfo{year}{2019}), \bibinfo{pages}{349--371}.
\newblock


\bibitem[Desai and Chin(2021)]%
        {DesaiChin2021}
\bibfield{author}{\bibinfo{person}{Smit Desai} {and} \bibinfo{person}{Jessie Chin}.} \bibinfo{year}{2021}\natexlab{}.
\newblock \showarticletitle{Hey Google, Can You Help Me Learn?}. In \bibinfo{booktitle}{\emph{Proceedings of the 3rd Conference on Conversational User Interfaces}} (Bilbao (online), Spain) \emph{(\bibinfo{series}{CUI '21})}. \bibinfo{publisher}{Association for Computing Machinery}, \bibinfo{address}{New York, NY, USA}, Article \bibinfo{articleno}{6}, \bibinfo{numpages}{4}~pages.
\newblock
\showISBNx{9781450389983}
\urldef\tempurl%
\url{https://doi.org/10.1145/3469595.3469601}
\showDOI{\tempurl}


\bibitem[Desai and Chin(2023)]%
        {DesaiChin2023}
\bibfield{author}{\bibinfo{person}{Smit Desai} {and} \bibinfo{person}{Jessie Chin}.} \bibinfo{year}{2023}\natexlab{}.
\newblock \showarticletitle{OK Google, Let's Learn: Using Voice User Interfaces for Informal Self-Regulated Learning of Health Topics among Younger and Older Adults}. In \bibinfo{booktitle}{\emph{Proceedings of the 2023 CHI Conference on Human Factors in Computing Systems}} (Hamburg, Germany) \emph{(\bibinfo{series}{CHI '23})}. \bibinfo{publisher}{Association for Computing Machinery}, \bibinfo{address}{New York, NY, USA}, Article \bibinfo{articleno}{847}, \bibinfo{numpages}{21}~pages.
\newblock
\showISBNx{9781450394215}
\urldef\tempurl%
\url{https://doi.org/10.1145/3544548.3581507}
\showDOI{\tempurl}


\bibitem[Desai et~al\mbox{.}(2024a)]%
        {Desai_Dubiel_Leiva_2024}
\bibfield{author}{\bibinfo{person}{Smit Desai}, \bibinfo{person}{Mateusz Dubiel}, {and} \bibinfo{person}{Luis~A. Leiva}.} \bibinfo{year}{2024}\natexlab{a}.
\newblock \showarticletitle{Examining Humanness as a Metaphor to Design Voice User Interfaces}. In \bibinfo{booktitle}{\emph{Proceedings of the 6th ACM Conference on Conversational User Interfaces}} \emph{(\bibinfo{series}{CUI ’24})}. \bibinfo{publisher}{Association for Computing Machinery}, \bibinfo{address}{New York, NY, USA}, \bibinfo{pages}{1–15}.
\newblock
\showISBNx{9798400705113}
\urldef\tempurl%
\url{https://doi.org/10.1145/3640794.3665535}
\showDOI{\tempurl}


\bibitem[Desai and Twidale(2022)]%
        {Desai_Twidale_2022}
\bibfield{author}{\bibinfo{person}{Smit Desai} {and} \bibinfo{person}{Michael Twidale}.} \bibinfo{year}{2022}\natexlab{}.
\newblock \showarticletitle{Is Alexa like a computer? A search engine? A friend? A silly child? Yes.}. In \bibinfo{booktitle}{\emph{Proceedings of the 4th Conference on Conversational User Interfaces}} (Glasgow, United Kingdom) \emph{(\bibinfo{series}{CUI '22})}. \bibinfo{publisher}{Association for Computing Machinery}, \bibinfo{address}{New York, NY, USA}, Article \bibinfo{articleno}{21}, \bibinfo{numpages}{4}~pages.
\newblock
\showISBNx{9781450397391}
\urldef\tempurl%
\url{https://doi.org/10.1145/3543829.3544535}
\showDOI{\tempurl}


\bibitem[Desai and Twidale(2023)]%
        {Desai_Twidale_2023}
\bibfield{author}{\bibinfo{person}{Smit Desai} {and} \bibinfo{person}{Michael Twidale}.} \bibinfo{year}{2023}\natexlab{}.
\newblock \showarticletitle{Metaphors in Voice User Interfaces: A Slippery Fish}.
\newblock \bibinfo{journal}{\emph{ACM Transactions on Computer-Human Interaction}} \bibinfo{volume}{30}, \bibinfo{number}{6} (\bibinfo{date}{Sept.} \bibinfo{year}{2023}), \bibinfo{pages}{89:1--89:37}.
\newblock
\showISSN{1073-0516}
\urldef\tempurl%
\url{https://doi.org/10.1145/3609326}
\showDOI{\tempurl}


\bibitem[Desai et~al\mbox{.}(2024b)]%
        {Desai2024CUICHI}
\bibfield{author}{\bibinfo{person}{Smit Desai}, \bibinfo{person}{Christina~Ziying Wei}, \bibinfo{person}{Jaisie Sin}, \bibinfo{person}{Mateusz Dubiel}, \bibinfo{person}{Nima Zargham}, \bibinfo{person}{Shashank Ahire}, \bibinfo{person}{Martin Porcheron}, \bibinfo{person}{Anastasia Kuzminykh}, \bibinfo{person}{Minha Lee}, \bibinfo{person}{Heloisa Candello}, \bibinfo{person}{Joel~E Fischer}, \bibinfo{person}{Cosmin Munteanu}, {and} \bibinfo{person}{Benjamin~R. Cowan}.} \bibinfo{year}{2024}\natexlab{b}.
\newblock \showarticletitle{CUI@CHI 2024: Building Trust in CUIs—From Design to Deployment}. In \bibinfo{booktitle}{\emph{Extended Abstracts of the 2024 CHI Conference on Human Factors in Computing Systems}} \emph{(\bibinfo{series}{CHI EA '24})}. \bibinfo{publisher}{Association for Computing Machinery}, \bibinfo{address}{New York, NY, USA}, Article \bibinfo{articleno}{460}, \bibinfo{numpages}{7}~pages.
\newblock
\showISBNx{9798400703317}
\urldef\tempurl%
\url{https://doi.org/10.1145/3613905.3636287}
\showDOI{\tempurl}


\bibitem[Doyle et~al\mbox{.}(2019)]%
        {Doyle2019Mapping}
\bibfield{author}{\bibinfo{person}{Philip~R. Doyle}, \bibinfo{person}{Justin Edwards}, \bibinfo{person}{Odile Dumbleton}, \bibinfo{person}{Leigh Clark}, {and} \bibinfo{person}{Benjamin~R. Cowan}.} \bibinfo{year}{2019}\natexlab{}.
\newblock \showarticletitle{Mapping Perceptions of Humanness in Intelligent Personal Assistant Interaction}. In \bibinfo{booktitle}{\emph{Proceedings of the 21st International Conference on Human-Computer Interaction with Mobile Devices and Services}} (Taipei, Taiwan) \emph{(\bibinfo{series}{MobileHCI '19})}. \bibinfo{publisher}{Association for Computing Machinery}, \bibinfo{address}{New York, NY, USA}, Article \bibinfo{articleno}{5}, \bibinfo{numpages}{12}~pages.
\newblock
\showISBNx{9781450368254}
\urldef\tempurl%
\url{https://doi.org/10.1145/3338286.3340116}
\showDOI{\tempurl}


\bibitem[Dubiel(2018)]%
        {dubiel2018towards}
\bibfield{author}{\bibinfo{person}{Mateusz Dubiel}.} \bibinfo{year}{2018}\natexlab{}.
\newblock \showarticletitle{Towards Human-Like Conversational Search Systems}. In \bibinfo{booktitle}{\emph{Proceedings of the 2018 Conference on Human Information Interaction \& Retrieval}} (New Brunswick, NJ, USA) \emph{(\bibinfo{series}{CHIIR '18})}. \bibinfo{publisher}{Association for Computing Machinery}, \bibinfo{address}{New York, NY, USA}, \bibinfo{pages}{348–350}.
\newblock
\showISBNx{9781450349253}
\urldef\tempurl%
\url{https://doi.org/10.1145/3176349.3176360}
\showDOI{\tempurl}


\bibitem[Dubiel et~al\mbox{.}(2019)]%
        {dubiel2019inquisitive}
\bibfield{author}{\bibinfo{person}{Mateusz Dubiel}, \bibinfo{person}{Alessandra Cervone}, {and} \bibinfo{person}{Giuseppe Riccardi}.} \bibinfo{year}{2019}\natexlab{}.
\newblock \showarticletitle{Inquisitive mind: a conversational news companion}. In \bibinfo{booktitle}{\emph{Proceedings of the 1st International Conference on Conversational User Interfaces}} (Dublin, Ireland) \emph{(\bibinfo{series}{CUI '19})}. \bibinfo{publisher}{Association for Computing Machinery}, \bibinfo{address}{New York, NY, USA}, Article \bibinfo{articleno}{6}, \bibinfo{numpages}{3}~pages.
\newblock
\showISBNx{9781450371872}
\urldef\tempurl%
\url{https://doi.org/10.1145/3342775.3342802}
\showDOI{\tempurl}


\bibitem[Dubiel et~al\mbox{.}(2022)]%
        {dubiel2022conversational}
\bibfield{author}{\bibinfo{person}{Mateusz Dubiel}, \bibinfo{person}{Sylvain Daronnat}, {and} \bibinfo{person}{Luis~A. Leiva}.} \bibinfo{year}{2022}\natexlab{}.
\newblock \showarticletitle{Conversational Agents Trust Calibration: A User-Centred Perspective to Design}. In \bibinfo{booktitle}{\emph{Proceedings of the 4th Conference on Conversational User Interfaces}} (Glasgow, United Kingdom) \emph{(\bibinfo{series}{CUI '22})}. \bibinfo{publisher}{Association for Computing Machinery}, \bibinfo{address}{New York, NY, USA}, Article \bibinfo{articleno}{30}, \bibinfo{numpages}{6}~pages.
\newblock
\showISBNx{9781450397391}
\urldef\tempurl%
\url{https://doi.org/10.1145/3543829.3544518}
\showDOI{\tempurl}


\bibitem[Dubiel et~al\mbox{.}(2024a)]%
        {voiceCraft2024}
\bibfield{author}{\bibinfo{person}{Mateusz Dubiel}, \bibinfo{person}{Smit Desai}, \bibinfo{person}{Nima Zargham}, {and} \bibinfo{person}{Anuschka Schmitt}.} \bibinfo{year}{2024}\natexlab{a}.
\newblock \showarticletitle{Voicecraft: Designing Task-specific Voice Assistant Personas}. In \bibinfo{booktitle}{\emph{Proceedings of the 6th ACM Conference on Conversational User Interfaces}} (Luxembourg, Luxembourg) \emph{(\bibinfo{series}{CUI '24})}. \bibinfo{publisher}{Association for Computing Machinery}, \bibinfo{address}{New York, NY, USA}, Article \bibinfo{articleno}{66}, \bibinfo{numpages}{3}~pages.
\newblock
\showISBNx{9798400705113}
\urldef\tempurl%
\url{https://doi.org/10.1145/3640794.3670000}
\showDOI{\tempurl}


\bibitem[Dubiel et~al\mbox{.}(2020)]%
        {dubiel2020persuasive}
\bibfield{author}{\bibinfo{person}{Mateusz Dubiel}, \bibinfo{person}{Martin Halvey}, \bibinfo{person}{Pilar~Oplustil Gallegos}, {and} \bibinfo{person}{Simon King}.} \bibinfo{year}{2020}\natexlab{}.
\newblock \showarticletitle{Persuasive Synthetic Speech: Voice Perception and User Behaviour}. In \bibinfo{booktitle}{\emph{Proceedings of the 2nd Conference on Conversational User Interfaces}} (Bilbao, Spain) \emph{(\bibinfo{series}{CUI '20})}. \bibinfo{publisher}{Association for Computing Machinery}, \bibinfo{address}{New York, NY, USA}, Article \bibinfo{articleno}{6}, \bibinfo{numpages}{9}~pages.
\newblock
\showISBNx{9781450375443}
\urldef\tempurl%
\url{https://doi.org/10.1145/3405755.3406120}
\showDOI{\tempurl}


\bibitem[Dubiel et~al\mbox{.}(2024b)]%
        {dubiel2024hey}
\bibfield{author}{\bibinfo{person}{Mateusz Dubiel}, \bibinfo{person}{Luis~A. Leiva}, \bibinfo{person}{Kerstin Bongard-Blanchy}, {and} \bibinfo{person}{Anastasia Sergeeva}.} \bibinfo{year}{2024}\natexlab{b}.
\newblock \bibinfo{title}{“Hey Genie, You Got Me Thinking About My Menu Choices!”: Impact of Proactive Feedback on User Perception and Reflection in Decision-making Tasks}.
\newblock
\newblock
\showISSN{1073-0516}
\urldef\tempurl%
\url{https://doi.org/10.1145/3685274}
\showDOI{\tempurl}
\newblock
\shownote{Just Accepted}.


\bibitem[Gray et~al\mbox{.}(2024a)]%
        {gray_mobilizing_2024}
\bibfield{author}{\bibinfo{person}{Colin~M. Gray}, \bibinfo{person}{Johanna~T. Gunawan}, \bibinfo{person}{Ren\'{e} Sch\"{a}fer}, \bibinfo{person}{Nataliia Bielova}, \bibinfo{person}{Lorena Sanchez~Chamorro}, \bibinfo{person}{Katie Seaborn}, \bibinfo{person}{Thomas Mildner}, {and} \bibinfo{person}{Hauke Sandhaus}.} \bibinfo{year}{2024}\natexlab{a}.
\newblock \showarticletitle{Mobilizing Research and Regulatory Action on Dark Patterns and Deceptive Design Practices}. In \bibinfo{booktitle}{\emph{Extended Abstracts of the 2024 CHI Conference on Human Factors in Computing Systems}} \emph{(\bibinfo{series}{CHI EA '24})}. \bibinfo{publisher}{Association for Computing Machinery}, \bibinfo{address}{New York, NY, USA}, Article \bibinfo{articleno}{482}, \bibinfo{numpages}{6}~pages.
\newblock
\showISBNx{9798400703317}
\urldef\tempurl%
\url{https://doi.org/10.1145/3613905.3636310}
\showDOI{\tempurl}


\bibitem[Gray et~al\mbox{.}(2024b)]%
        {gray_ontology_2024}
\bibfield{author}{\bibinfo{person}{Colin~M. Gray}, \bibinfo{person}{Cristiana~Teixeira Santos}, \bibinfo{person}{Nataliia Bielova}, {and} \bibinfo{person}{Thomas Mildner}.} \bibinfo{year}{2024}\natexlab{b}.
\newblock \showarticletitle{An {Ontology} of {Dark} {Patterns} {Knowledge}: {Foundations}, {Definitions}, and a {Pathway} for {Shared} {Knowledge}-{Building}}. In \bibinfo{booktitle}{\emph{Proceedings of the {CHI} {Conference} on {Human} {Factors} in {Computing} {Systems}}}. \bibinfo{publisher}{ACM}, \bibinfo{address}{Honolulu HI USA}, \bibinfo{pages}{1--22}.
\newblock
\showISBNx{9798400703300}
\urldef\tempurl%
\url{https://doi.org/10.1145/3613904.3642436}
\showDOI{\tempurl}


\bibitem[Hwang et~al\mbox{.}(2024)]%
        {10.1145/3643834.3661555}
\bibfield{author}{\bibinfo{person}{Angel Hsing-Chi Hwang}, \bibinfo{person}{John~Oliver Siy}, \bibinfo{person}{Renee Shelby}, {and} \bibinfo{person}{Alison Lentz}.} \bibinfo{year}{2024}\natexlab{}.
\newblock \showarticletitle{In Whose Voice?: Examining AI Agent Representation of People in Social Interaction through Generative Speech}. In \bibinfo{booktitle}{\emph{Proceedings of the 2024 ACM Designing Interactive Systems Conference}} (Copenhagen, Denmark) \emph{(\bibinfo{series}{DIS '24})}. \bibinfo{publisher}{Association for Computing Machinery}, \bibinfo{address}{New York, NY, USA}, \bibinfo{pages}{224–245}.
\newblock
\showISBNx{9798400705830}
\urldef\tempurl%
\url{https://doi.org/10.1145/3643834.3661555}
\showDOI{\tempurl}


\bibitem[Jentsch et~al\mbox{.}(2019)]%
        {jentsch2019talking}
\bibfield{author}{\bibinfo{person}{Martin Jentsch}, \bibinfo{person}{Maresa Biermann}, {and} \bibinfo{person}{Evelyn Schweiger}.} \bibinfo{year}{2019}\natexlab{}.
\newblock \showarticletitle{Talking to Stupid?!? Improving Voice User Interfaces}. In \bibinfo{booktitle}{\emph{Mensch und Computer 2019 - Usability Professionals}}. \bibinfo{publisher}{Gesellschaft für Informatik e.V. Und German UPA e.V.}, \bibinfo{address}{Bonn}.
\newblock


\bibitem[Khan and De~Angeli(2009)]%
        {khan2009attractiveness}
\bibfield{author}{\bibinfo{person}{Rabia Khan} {and} \bibinfo{person}{Antonella De~Angeli}.} \bibinfo{year}{2009}\natexlab{}.
\newblock \showarticletitle{The attractiveness stereotype in the evaluation of embodied conversational agents}. In \bibinfo{booktitle}{\emph{IFIP Conference on Human-Computer Interaction}}. Springer, \bibinfo{publisher}{Springer}, \bibinfo{address}{Heidelberg, Germany}, \bibinfo{pages}{85--97}.
\newblock


\bibitem[Lopatovska et~al\mbox{.}(2019)]%
        {Lopatovska:2019}
\bibfield{author}{\bibinfo{person}{Irene Lopatovska}, \bibinfo{person}{Katrina Rink}, \bibinfo{person}{Ian Knight}, \bibinfo{person}{Kieran Raines}, \bibinfo{person}{Kevin Cosenza}, \bibinfo{person}{Harriet Williams}, \bibinfo{person}{Perachya Sorsche}, \bibinfo{person}{David Hirsch}, \bibinfo{person}{Qi Li}, {and} \bibinfo{person}{Adrianna Martinez}.} \bibinfo{year}{2019}\natexlab{}.
\newblock \showarticletitle{Talk to me: Exploring user interactions with the Amazon Alexa}.
\newblock \bibinfo{journal}{\emph{Journal of Librarianship and Information Science}} \bibinfo{volume}{51}, \bibinfo{number}{4} (\bibinfo{year}{2019}), \bibinfo{pages}{984--997}.
\newblock
\urldef\tempurl%
\url{https://doi.org/10.1177/0961000618759414}
\showDOI{\tempurl}
\showeprint{https://doi.org/10.1177/0961000618759414}


\bibitem[Luger and Sellen(2016)]%
        {Luger:2016}
\bibfield{author}{\bibinfo{person}{Ewa Luger} {and} \bibinfo{person}{Abigail Sellen}.} \bibinfo{year}{2016}\natexlab{}.
\newblock \showarticletitle{"Like Having a Really Bad PA": The Gulf Between User Expectation and Experience of Conversational Agents}. In \bibinfo{booktitle}{\emph{Proceedings of the 2016 CHI Conference on Human Factors in Computing Systems}} (San Jose, California, USA) \emph{(\bibinfo{series}{CHI '16})}. \bibinfo{publisher}{ACM}, \bibinfo{address}{New York, NY, USA}, \bibinfo{pages}{5286--5297}.
\newblock
\showISBNx{978-1-4503-3362-7}
\urldef\tempurl%
\url{https://doi.org/10.1145/2858036.2858288}
\showDOI{\tempurl}


\bibitem[McMillan et~al\mbox{.}(2023)]%
        {McMillan2023}
\bibfield{author}{\bibinfo{person}{Donald McMillan}, \bibinfo{person}{Razan Jaber}, \bibinfo{person}{Benjamin~R. Cowan}, \bibinfo{person}{Joel~E. Fischer}, \bibinfo{person}{Bahar Irfan}, \bibinfo{person}{Ronald Cumbal}, \bibinfo{person}{Nima Zargham}, {and} \bibinfo{person}{Minha Lee}.} \bibinfo{year}{2023}\natexlab{}.
\newblock \showarticletitle{Human-Robot Conversational Interaction (HRCI)}. In \bibinfo{booktitle}{\emph{Companion of the 2023 ACM/IEEE International Conference on Human-Robot Interaction}} (Stockholm, Sweden) \emph{(\bibinfo{series}{HRI '23})}. \bibinfo{publisher}{Association for Computing Machinery}, \bibinfo{address}{New York, NY, USA}, \bibinfo{pages}{923–925}.
\newblock
\showISBNx{9781450399708}
\urldef\tempurl%
\url{https://doi.org/10.1145/3568294.3579954}
\showDOI{\tempurl}


\bibitem[Meta(2023)]%
        {Introducing2023}
\bibfield{author}{\bibinfo{person}{Meta}.} \bibinfo{year}{2023}\natexlab{}.
\newblock \bibinfo{title}{Introducing New AI Experiences Across Our Family of Apps and Devices}.
\newblock
\newblock
\urldef\tempurl%
\url{https://about.fb.com/news/2023/09/introducing-ai-powered-assistants-characters-and-creative-tools/}
\showURL{%
\tempurl}


\bibitem[Mildner et~al\mbox{.}(2024)]%
        {mildner_listening_2024}
\bibfield{author}{\bibinfo{person}{Thomas Mildner}, \bibinfo{person}{Orla Cooney}, \bibinfo{person}{Anna-Maria Meck}, \bibinfo{person}{Marion Bartl}, \bibinfo{person}{Gian-Luca Savino}, \bibinfo{person}{Philip~R Doyle}, \bibinfo{person}{Diego Garaialde}, \bibinfo{person}{Leigh Clark}, \bibinfo{person}{John Sloan}, \bibinfo{person}{Nina Wenig}, \bibinfo{person}{Rainer Malaka}, {and} \bibinfo{person}{Jasmin Niess}.} \bibinfo{year}{2024}\natexlab{}.
\newblock \showarticletitle{Listening to the {Voices}: {Describing} {Ethical} {Caveats} of {Conversational} {User} {Interfaces} {According} to {Experts} and {Frequent} {Users}}. In \bibinfo{booktitle}{\emph{Proceedings of the {CHI} {Conference} on {Human} {Factors} in {Computing} {Systems}}}. \bibinfo{publisher}{ACM}, \bibinfo{address}{Honolulu HI USA}, \bibinfo{pages}{1--18}.
\newblock
\showISBNx{9798400703300}
\urldef\tempurl%
\url{https://doi.org/10.1145/3613904.3642542}
\showDOI{\tempurl}


\bibitem[Mildner et~al\mbox{.}(2022)]%
        {mildner_rules_2022}
\bibfield{author}{\bibinfo{person}{Thomas Mildner}, \bibinfo{person}{Philip Doyle}, \bibinfo{person}{Gian-Luca Savino}, {and} \bibinfo{person}{Rainer Malaka}.} \bibinfo{year}{2022}\natexlab{}.
\newblock \showarticletitle{Rules {Of} {Engagement}: {Levelling} {Up} {To} {Combat} {Unethical} {CUI} {Design}}. In \bibinfo{booktitle}{\emph{4th {Conference} on {Conversational} {User} {Interfaces}}}. \bibinfo{publisher}{ACM}, \bibinfo{address}{Glasgow United Kingdom}, \bibinfo{pages}{1--5}.
\newblock
\showISBNx{978-1-4503-9739-1}
\urldef\tempurl%
\url{https://doi.org/10.1145/3543829.3544528}
\showDOI{\tempurl}


\bibitem[Moore and Arar(2019)]%
        {Moore2019Conversational}
\bibfield{author}{\bibinfo{person}{Robert~J. Moore} {and} \bibinfo{person}{Raphael Arar}.} \bibinfo{year}{2019}\natexlab{}.
\newblock \bibinfo{booktitle}{\emph{Conversational UX Design: A Practitioner's Guide to the Natural Conversation Framework}}.
\newblock \bibinfo{publisher}{Association for Computing Machinery}, \bibinfo{address}{New York, NY, USA}.
\newblock
\showISBNx{9781450363013}
\urldef\tempurl%
\url{https://doi.org/10.1145/3304087}
\showDOI{\tempurl}


\bibitem[Motalebi et~al\mbox{.}(2019)]%
        {MotalebiChoSundarAbdullah2019}
\bibfield{author}{\bibinfo{person}{Nasim Motalebi}, \bibinfo{person}{Eugene Cho}, \bibinfo{person}{S.~Shyam Sundar}, {and} \bibinfo{person}{Saeed Abdullah}.} \bibinfo{year}{2019}\natexlab{}.
\newblock \showarticletitle{Can Alexa be your Therapist? How Back-Channeling Transforms Smart-Speakers to be Active Listeners}. In \bibinfo{booktitle}{\emph{Companion Publication of the 2019 Conference on Computer Supported Cooperative Work and Social Computing}} (Austin, TX, USA) \emph{(\bibinfo{series}{CSCW '19 Companion})}. \bibinfo{publisher}{Association for Computing Machinery}, \bibinfo{address}{New York, NY, USA}, \bibinfo{pages}{309–313}.
\newblock
\showISBNx{9781450366922}
\urldef\tempurl%
\url{https://doi.org/10.1145/3311957.3359502}
\showDOI{\tempurl}


\bibitem[Murad and Munteanu(2019)]%
        {murad2019don}
\bibfield{author}{\bibinfo{person}{Christine Murad} {and} \bibinfo{person}{Cosmin Munteanu}.} \bibinfo{year}{2019}\natexlab{}.
\newblock \showarticletitle{"I Don't Know What You're Talking about, HALexa": The Case for Voice User Interface Guidelines}. In \bibinfo{booktitle}{\emph{Proceedings of the 1st International Conference on Conversational User Interfaces}} (Dublin, Ireland) \emph{(\bibinfo{series}{CUI '19})}. \bibinfo{publisher}{Association for Computing Machinery}, \bibinfo{address}{New York, NY, USA}, Article \bibinfo{articleno}{9}, \bibinfo{numpages}{3}~pages.
\newblock
\showISBNx{9781450371872}
\urldef\tempurl%
\url{https://doi.org/10.1145/3342775.3342795}
\showDOI{\tempurl}


\bibitem[Reicherts et~al\mbox{.}(2021)]%
        {reicherts2021MayI}
\bibfield{author}{\bibinfo{person}{Leon Reicherts}, \bibinfo{person}{Nima Zargham}, \bibinfo{person}{Michael Bonfert}, \bibinfo{person}{Yvonne Rogers}, {and} \bibinfo{person}{Rainer Malaka}.} \bibinfo{year}{2021}\natexlab{}.
\newblock \showarticletitle{May I Interrupt? Diverging Opinions On Proactive Smart Speakers}. In \bibinfo{booktitle}{\emph{Proceedings of the 3rd Conference on Conversational User Interfaces}} (Bilbao (online), Spain) \emph{(\bibinfo{series}{CUI '21})}. \bibinfo{publisher}{Association for Computing Machinery}, \bibinfo{address}{New York, NY, USA}, Article \bibinfo{articleno}{34}, \bibinfo{numpages}{10}~pages.
\newblock
\showISBNx{9781450389983}
\urldef\tempurl%
\url{https://doi.org/10.1145/3469595.3469629}
\showDOI{\tempurl}


\bibitem[Ruan et~al\mbox{.}(2023)]%
        {ruan2023tptu}
\bibfield{author}{\bibinfo{person}{Jingqing Ruan}, \bibinfo{person}{Yihong Chen}, \bibinfo{person}{Bin Zhang}, \bibinfo{person}{Zhiwei Xu}, \bibinfo{person}{Tianpeng Bao}, \bibinfo{person}{Guoqing Du}, \bibinfo{person}{Shiwei Shi}, \bibinfo{person}{Hangyu Mao}, \bibinfo{person}{Xingyu Zeng}, {and} \bibinfo{person}{Rui Zhao}.} \bibinfo{year}{2023}\natexlab{}.
\newblock \bibinfo{title}{TPTU: Task Planning and Tool Usage of Large Language Model-based AI Agents}.
\newblock , \bibinfo{numpages}{arXiv--2308}~pages.
\newblock
\urldef\tempurl%
\url{https://openreview.net/forum?id=GrkgKtOjaH}
\showURL{%
\tempurl}


\bibitem[Schmitt et~al\mbox{.}(2021)]%
        {schmitt2021voice}
\bibfield{author}{\bibinfo{person}{Anuschka Schmitt}, \bibinfo{person}{Naim Zierau}, \bibinfo{person}{Andreas Janson}, {and} \bibinfo{person}{Jan~Marco Leimeister}.} \bibinfo{year}{2021}\natexlab{}.
\newblock \bibinfo{title}{Voice as a contemporary frontier of interaction design}.
\newblock
\newblock


\bibitem[Trajkova and Martin-Hammond(2020)]%
        {Trajkova_Martin-Hammond_2020}
\bibfield{author}{\bibinfo{person}{Milka Trajkova} {and} \bibinfo{person}{Aqueasha Martin-Hammond}.} \bibinfo{year}{2020}\natexlab{}.
\newblock \bibinfo{booktitle}{\emph{“Alexa is a Toy”: Exploring Older Adults’ Reasons for Using, Limiting, and Abandoning Echo}}.
\newblock \bibinfo{publisher}{Association for Computing Machinery}, \bibinfo{address}{New York, NY, USA}, \bibinfo{pages}{1–13}.
\newblock
\showISBNx{978-1-4503-6708-0}
\urldef\tempurl%
\url{https://doi.org/10.1145/3313831.3376760}
\showURL{%
\tempurl}


\bibitem[Wang and Ruiz(2021)]%
        {Wang2021Examining}
\bibfield{author}{\bibinfo{person}{Isaac Wang} {and} \bibinfo{person}{J. Ruiz}.} \bibinfo{year}{2021}\natexlab{}.
\newblock \showarticletitle{Examining the Use of Nonverbal Communication in Virtual Agents}.
\newblock \bibinfo{journal}{\emph{International Journal of Human–Computer Interaction}}  \bibinfo{volume}{37} (\bibinfo{year}{2021}), \bibinfo{pages}{1648 -- 1673}.
\newblock
\urldef\tempurl%
\url{https://doi.org/10.1080/10447318.2021.1898851}
\showDOI{\tempurl}


\bibitem[Yuksel et~al\mbox{.}(2017)]%
        {Yuksel:2017}
\bibfield{author}{\bibinfo{person}{Beste~F. Yuksel}, \bibinfo{person}{Penny Collisson}, {and} \bibinfo{person}{Mary Czerwinski}.} \bibinfo{year}{2017}\natexlab{}.
\newblock \showarticletitle{Brains or Beauty: How to Engender Trust in User-Agent Interactions}.
\newblock \bibinfo{journal}{\emph{ACM Trans. Internet Technol.}} \bibinfo{volume}{17}, \bibinfo{number}{1}, Article \bibinfo{articleno}{2} (\bibinfo{date}{Jan.} \bibinfo{year}{2017}), \bibinfo{numpages}{20}~pages.
\newblock
\showISSN{1533-5399}
\urldef\tempurl%
\url{https://doi.org/10.1145/2998572}
\showDOI{\tempurl}


\bibitem[Zargham(2024)]%
        {ZarghamDissertation2024}
\bibfield{author}{\bibinfo{person}{Nima Zargham}.} \bibinfo{year}{2024}\natexlab{}.
\newblock \emph{\bibinfo{title}{Expanding Speech Interaction for Domestic Activities}}.
\newblock \bibinfo{thesistype}{Ph.\,D. Dissertation}. \bibinfo{school}{University of Bremen}.
\newblock


\bibitem[Zargham et~al\mbox{.}(2022a)]%
        {zargham2021Customization}
\bibfield{author}{\bibinfo{person}{Nima Zargham}, \bibinfo{person}{Dmitry Alexandrovsky}, \bibinfo{person}{Jan Erich}, \bibinfo{person}{Nina Wenig}, {and} \bibinfo{person}{Rainer Malaka}.} \bibinfo{year}{2022}\natexlab{a}.
\newblock \showarticletitle{“I Want It That Way”: Exploring Users’ Customization and Personalization Preferences for Home Assistants}. In \bibinfo{booktitle}{\emph{Extended Abstracts of the 2022 CHI Conference on Human Factors in Computing Systems}} (New Orleans, LA, USA) \emph{(\bibinfo{series}{CHI EA '22})}. \bibinfo{publisher}{Association for Computing Machinery}, \bibinfo{address}{New York, NY, USA}, Article \bibinfo{articleno}{270}, \bibinfo{numpages}{8}~pages.
\newblock
\showISBNx{9781450391566}
\urldef\tempurl%
\url{https://doi.org/10.1145/3491101.3519843}
\showDOI{\tempurl}


\bibitem[Zargham et~al\mbox{.}(2023)]%
        {Zargham2023FaceIt}
\bibfield{author}{\bibinfo{person}{Nima Zargham}, \bibinfo{person}{Dmitry Alexandrovsky}, \bibinfo{person}{Thomas Mildner}, \bibinfo{person}{Robert Porzel}, {and} \bibinfo{person}{Rainer Malaka}.} \bibinfo{year}{2023}\natexlab{}.
\newblock \showarticletitle{"Let's Face It": Investigating User Preferences for Virtual Humanoid Home Assistants}. In \bibinfo{booktitle}{\emph{Proceedings of the 11th International Conference on Human-Agent Interaction}} (Gothenburg, Sweden) \emph{(\bibinfo{series}{HAI '23})}. \bibinfo{publisher}{Association for Computing Machinery}, \bibinfo{address}{New York, NY, USA}, \bibinfo{pages}{246–256}.
\newblock
\showISBNx{9798400708244}
\urldef\tempurl%
\url{https://doi.org/10.1145/3623809.3623821}
\showDOI{\tempurl}


\bibitem[Zargham et~al\mbox{.}(2021)]%
        {zargham2021MultiAgent}
\bibfield{author}{\bibinfo{person}{Nima Zargham}, \bibinfo{person}{Michael Bonfert}, \bibinfo{person}{Robert Porzel}, \bibinfo{person}{Tanja Doring}, {and} \bibinfo{person}{Rainer Malaka}.} \bibinfo{year}{2021}\natexlab{}.
\newblock \showarticletitle{Multi-Agent Voice Assistants: An Investigation Of User Experience}. In \bibinfo{booktitle}{\emph{20th International Conference on Mobile and Ubiquitous Multimedia}} (Leuven, Belgium) \emph{(\bibinfo{series}{MUM 2021})}. \bibinfo{publisher}{Association for Computing Machinery}, \bibinfo{address}{New York, NY, USA}, \bibinfo{pages}{98–107}.
\newblock
\showISBNx{9781450386432}
\urldef\tempurl%
\url{https://doi.org/10.1145/3490632.3490662}
\showDOI{\tempurl}


\bibitem[Zargham et~al\mbox{.}(2022b)]%
        {zargham2022Proactive}
\bibfield{author}{\bibinfo{person}{Nima Zargham}, \bibinfo{person}{Leon Reicherts}, \bibinfo{person}{Michael Bonfert}, \bibinfo{person}{Sarah~Theres Voelkel}, \bibinfo{person}{Johannes Schoening}, \bibinfo{person}{Rainer Malaka}, {and} \bibinfo{person}{Yvonne Rogers}.} \bibinfo{year}{2022}\natexlab{b}.
\newblock \showarticletitle{Understanding Circumstances for Desirable Proactive Behaviour of Voice Assistants: The Proactivity Dilemma}. In \bibinfo{booktitle}{\emph{Proceedings of the 4th Conference on Conversational User Interfaces}} (Glasgow, United Kingdom) \emph{(\bibinfo{series}{CUI '22})}. \bibinfo{publisher}{Association for Computing Machinery}, \bibinfo{address}{New York, NY, USA}, Article \bibinfo{articleno}{3}, \bibinfo{numpages}{14}~pages.
\newblock
\showISBNx{9781450397391}
\urldef\tempurl%
\url{https://doi.org/10.1145/3543829.3543834}
\showDOI{\tempurl}


\end{thebibliography}

\end{document}